\documentclass[useAMS,usenatbib]{mn2e}
\usepackage{graphicx,graphics,amsmath}
\title[Observations of GX 304-1 during an Outburst]{Timing and spectral studies of the transient X-ray pulsar GX 304-1 during an outburst}
\author[Jincy Devasia, Marykutty James, Biswajit Paul and Kavila Indulekha]{Jincy Devasia$^{1,2}$\thanks{E-mail: jincydevasia@yahoo.com}, Marykutty James$^{1,2}$\thanks{E-mail:marykuttykjames@yahoo.co.in}, Biswajit Paul$^{2}$,  and Kavila Indulekha$^{1}$\\
$^{1}$School of Pure and Applied Physics, Mahatma Gandhi University, Kottayam-686560, Kerala, India\\
$^{2}$Raman Research Institute, Sadashivanagar, C. V. Raman Avenue, Bangalore 560080, India}

\begin{document}
\maketitle
\begin{abstract}We present the timing and spectral properties of the transient X-ray pulsar GX 304-1 during its recent outburst in 2010 August, using observations carried out with the Proportional Counter Array (PCA) instrument on-board the {\em Rossi X-ray Timing Explorer (RXTE)} satellite. We detected strong intensity and energy dependent variations in the pulse profiles during the outburst. The pulse profile showed significant evolution over the outburst. It showed complex structures consisting of a main peak with steps on both sides during the start of the outburst. On some days, a sharp dip like feature was seen which disappeared at the end of the outburst; when the profile evolved into a sinusoidal shape. At low energies, the pulse profiles appeared complex, consisting of multiple peaks and a narrow minimum. The amplitude of the second brightest peak in low energies decreased with energy, and above 12 keV, the shape of the pulse profile changed to a single broad peak with a dip like feature. The dip had energy dependence, both in phase and in width. We detected Quasi-periodic Oscillations (QPO) at 0.125 Hz with a harmonic. The QPO feature had a low  rms value of 2.9\% and it showed a positive energy dependence up to 40 keV with the rms value increasing to 9\% at 40 keV. The QPO frequency decreased from 0.128 Hz to 0.108 Hz in 12 days. During most of the outburst, the 3-30 keV spectrum of GX 304-1 can be well fitted with a partial covering power-law model with a high energy cut-off and iron fluorescent line emission. For a few of the observations carried out during the decay of the outburst, the partial covering absorption component is found to change to single component absorption. We also found that the partial covering and high energy cut-off parameters vary significantly with the pulse phase.
\end{abstract}
\begin{keywords}
 binaries: general - pulsars: individual: GX 304-1 - X-rays binaries - X-rays: individual: GX 304-1 - X-rays: stars
\end{keywords}
\section{INTRODUCTION}
The transient High Mass X-ray Binary pulsar GX 304-1 (4U 1258-61) was discovered in a balloon observation in 1967. Subsequent observations showed a large variability of the X-ray flux of GX 304-1 along with short bursts lasting for an hour \citep{Lewin1968a, Lewin1968b, Ricker1973}. X-ray pulsations with a period of $\sim$272 s were detected from {\em SAS-3} observations \citep{McClintock1977}. Pulsations were detected up to about 10 keV and the pulse profile was characterised by a single broad peak along with a weak second peak. The spectrum of GX 304-1 was found to be relatively hard with a photon index ($\Gamma$) of 1.5 $\pm$ 0.2. An important feature of this source is the presence of strong flares with a duration of $\sim$100 s \citep{McClintock1977}.    

The optical counterpart of this source was discovered to be a {\it m}$_{v}$ = 15 shell star with strong double peaked H$_\alpha$ emission and spectral type B2 Vne \citep{Parkes1980}. The mean X-ray luminosity of GX 304-1 was measured to be $\sim$ 10$^{35}$ ergs s$^{-1}$ at d=2$\pm$1 kpc. The source was thus added to a list of low-luminosity slowly-pulsating X-ray sources associated with Be stars \citep{Mason1978}. The remarkably large, blue shifted O$_I$ $\lambda$ 7773 absorption feature in the counterpart of GX 304-1 confirms its identification as a shell star, a subclass of Be stars \citep{Thomas1979}, for which a considerable amount of ejected mass forms an equatorially expanding shell around the star.

 Long term behaviour of this source, studied using data obtained from the {\em Vela} satellite revealed a 132.5 day periodicity \citep{P&T1983}. In the absence of any other periodicity in this source, the 132.5 day feature was interpreted as the orbital period of the neutron star, in an eccentric orbit with the Be star companion. In the year 1986, {\em EXOSAT} observations were made, during the time an outburst was expected from this source. The flux measured in the low (0.03-2.4 keV) and medium (2-6 keV) energy range were both 3 $\times$ 10$^{-12}$ erg cm $^{-2}$ s$^{-1}$ only. This was a factor of 25 below the quiescent flux level reported before 1980 using other satellites, and it was concluded that the source was in an "off state". Follow up studies of the low X-ray activity of this source with optical observations attributed the decreased activity to a drastic change in the shell structure of the companion star. Long term optical spectroscopic and photometric studies carried out in the 1980s indicated that the optical companion is in an inactive state having apparently lost part of its envelope \citep{Corbet1985, Pietsch1986, Haefner1988}. 

After the extended low activity from since the early 1980s, ISIS/ISGRI on board {\em International Gamma-Ray Astrophysics Laboratory} ({\em INTEGRAL}) detected the source GX 304-1 above 20 keV in 2008 June/July. They could see a softening of the spectrum in the 18 - 60 keV range as the source had brightened with a flux level of 3-7 mCrab \citep{Manousakis2008}. Again in 2009, {\em RXTE}-ASM and {\em Swift}-BAT light curve showed an outburst, peaking around July 10 and subsequently on November 14, {\em MAXI}/GSC detected an outburst which had started around November 11 with a daily average flux in the range of 25-54 mCrab. Correlating this outburst with {\em RXTE} and {\em Swift} observations, it was found that the start of this outburst is consistent with the recurrent period of 132.5 days which indicates that GX 304-1 is again in its active phase \citep{Yamamoto2009}. Later in March 2010, {\em Swift}-BAT detected a relatively long sustained outburst in which the source flux rose from 8 mCrab to 160 mCrab in the 15-50 keV band within 15 days \citep{Krimm2010}. In 2010 August 6, {\em MAXI}-GSC detected another recurrent outburst in the 4-10 keV band (58 mCrab). They noticed that the peak flux of this source has been gradually increasing  in the recent several successive outbursts. During this last known outburst, the source flux peaked at about 700 mCrab and {\em Suzaku} observations revealed a cyclotron absorption line at 51 keV, confirming the high magnetic field strength of the neutron star \citep{Mihara2010, Sakamoto2010}. 

\begin{figure}
\centering
\includegraphics[width=8 cm, angle=-90]{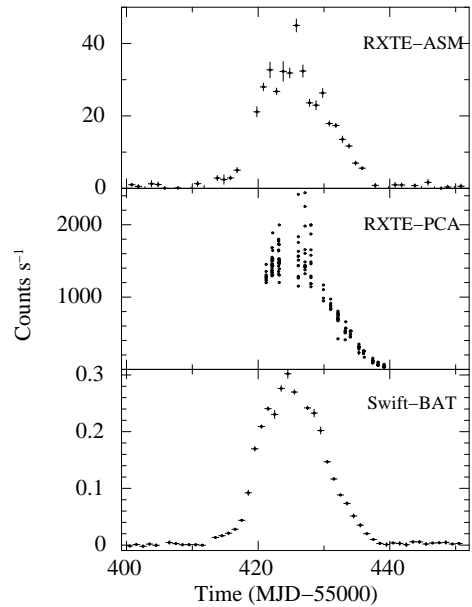}
\caption{The top panel shows the {\em RXTE}-ASM light curve of GX 304-1 in the 2-15 keV energy band with one day bin size, the middle panel shows the PCU2 light curve with a bin size same as spin period and the bottom panel shows the 15-50 keV light curve from the {\em Swift}-BAT all sky monitor with a one day bin size.}
\label{asm-pcu}
\end{figure}

In this work, we present the results of timing and spectral studies of GX 304-1 during the 2010 August outburst using data obtained from {\em RXTE} satellite. Pulse profiles of X-ray pulsars help us in understanding the pulsar geometry and beaming patterns. Most of the accretion powered X-ray pulsars exhibit strong intensity and energy dependence of pulse profiles and studies related to this provide insight regarding the relative contributions from the matter along the accretion column as well as from the scattering atmospheres formed surrounding the column. Aperiodic variabilities like quasi-periodic oscillations (QPOs) are seen in several transient as well as persistent High Mass X-ray Binary (HMXB) pulsars that are thought to arise due to the inhomogeneities in the inner accretion disk$-$ mainly the region where the accretion disk interacts with the pulsar magnetosphere. Studies on QPOs at different intensity levels give us information regarding the interaction of the inner accretion disc with the central object. In section 2, we describe the data and observations used for the present analysis. Section 3 details the timing analysis, mainly the periodic and aperiodic time variability of GX 304-1. Section 4 deals with the pulse phase averaged and resolved spectral analysis followed by a discussion of the results of the analysis in section 5.

\section{OBSERVATIONS AND DATA}
Several outbursts have been detected from the HMXB transient pulsar GX 304-1 after a 28 year quiescence. The 2010 August outburst was the sixth of  several successive outbursts with increased count rate compared to those before \citep {Mihara2010}. The {\em RXTE} targeted this source for a period covering almost the whole of this outburst starting from MJD 55421 to MJD 55439 (13th to 31st August 2010). {\em RXTE} carries three instruments: an All Sky Monitor (ASM) sensitive in the 1.5-12 keV energy range scanning the X-ray sky every $\sim$90 minutes and two non-imaging instruments: the High Energy X-ray Timing Experiment (HEXTE) comprising of two NaI/CsI scintillation detectors sensitive to X-ray photons in the 15-250 keV energy range, and the Proportional Counter Array (PCA) consisting of five identical, co-aligned proportional counter units (PCUs) sensitive in the 2-60 keV energy band with a total effective area of 6500 cm$^{2}$. {\em RXTE} had 23 pointings towards this source with ObsID P95417 for a total exposure of $\sim$ 62 ks. We have used data products obtained from the {\em RXTE}-PCA instrument for all the analyses mentioned in this paper. During most of the PCA observations, only one or two PCUs were ON. For the present analysis, we have used data only from the PCU2 detector which was ON for most of the time. The long term light curves of GX 304-1 are shown in Figure~\ref{asm-pcu} for different energy bands. The top panel shows the ASM light curve $(xte.mit.edu/ASM_lc.html)$ in 2-15 keV energy band with one day bin size, the middle panel shows the PCU2 light curve with a bin size same as the spin period and the bottom panel shows the 15-50 keV light curve from the {\em Swift}-BAT all sky monitor $(heasarc.gsfc.nasa.gov/docs/swift/results/transients/)$ with a one day bin size. We have used generic event and binned mode data and Standard1 mode data for creating the light curves and for all the following pulsation analysis. For the spectral analysis, we have used Standard2 mode data which provide 129 PHA channels covering the 2-60 keV energy range with 16 s time resolution. The extracted source spectra in the 3$-$30 keV range were analysed by fitting appropriate models using {\em XSPEC} v12.6.0 analysis package and a systematic error of 0.5\% is used to account for the calibration uncertainties.   

\section{Timing Analysis}
 \subsection{Light curves and pulse profiles}

\begin{figure}
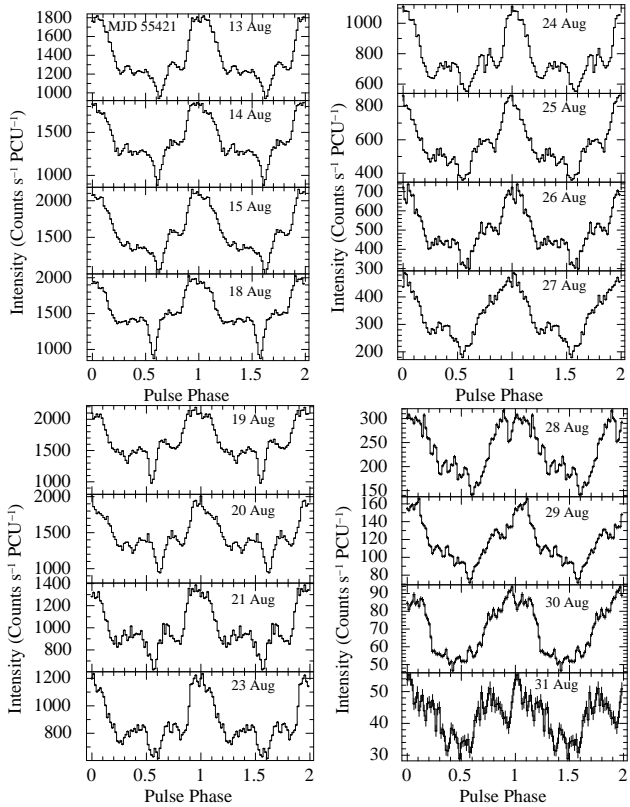

\centering
\includegraphics[width=5.3 cm, angle=-90]{fig2_a.ps}
\includegraphics[width=5.3 cm, angle=-90]{fig2_c.ps}
\includegraphics[width=5.3 cm, angle=-90]{fig2_b.ps}
\includegraphics[width=5.3 cm, angle=-90]{fig2_d.ps}
\caption{Pulse profiles (in 2-60 keV) of GX 304-1 on different days starting from Aug 13 (MJD 55421) to Aug 31 (MJD 55439) observed with {\em RXTE} during the outburst. The alignment of the pulse profiles for different days are done by hand, to match the peak.}
\label{fig2}
\end{figure}

\begin{figure}
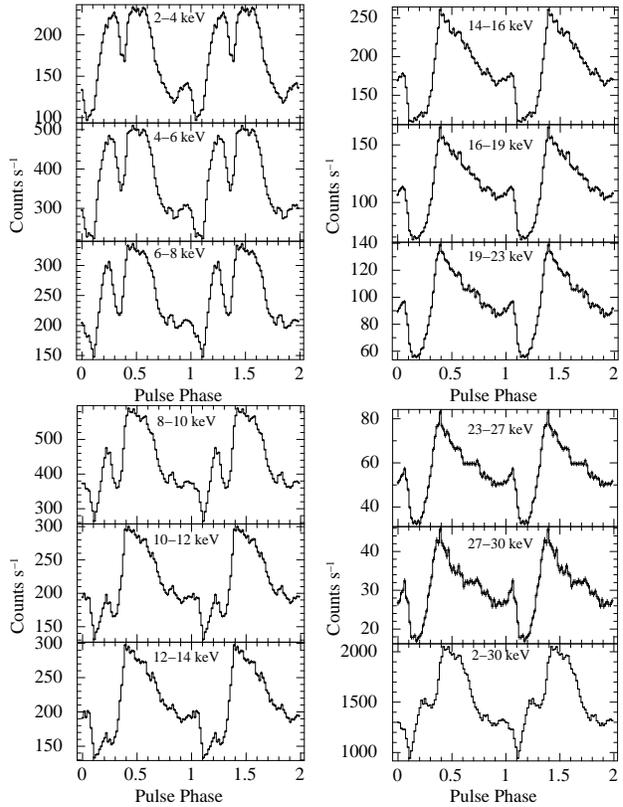

\centering
\includegraphics[width=5.3 cm, angle=-90]{fig3_a.ps}
\includegraphics[width=5.3 cm, angle=-90]{fig3_c.ps}
\includegraphics[width=5.3 cm, angle=-90]{fig3_b.ps}
\includegraphics[width=5.3 cm, angle=-90]{fig3_d.ps}
\caption{Energy dependent pulse profiles of GX 304-1 in different energy bands observed with {\em RXTE} during the peak phase of the outburst (August 15$-$ MJD 55423)}
\label{fig3}
\end{figure}

 \begin{figure*}
\centering
\includegraphics[width=18 cm]{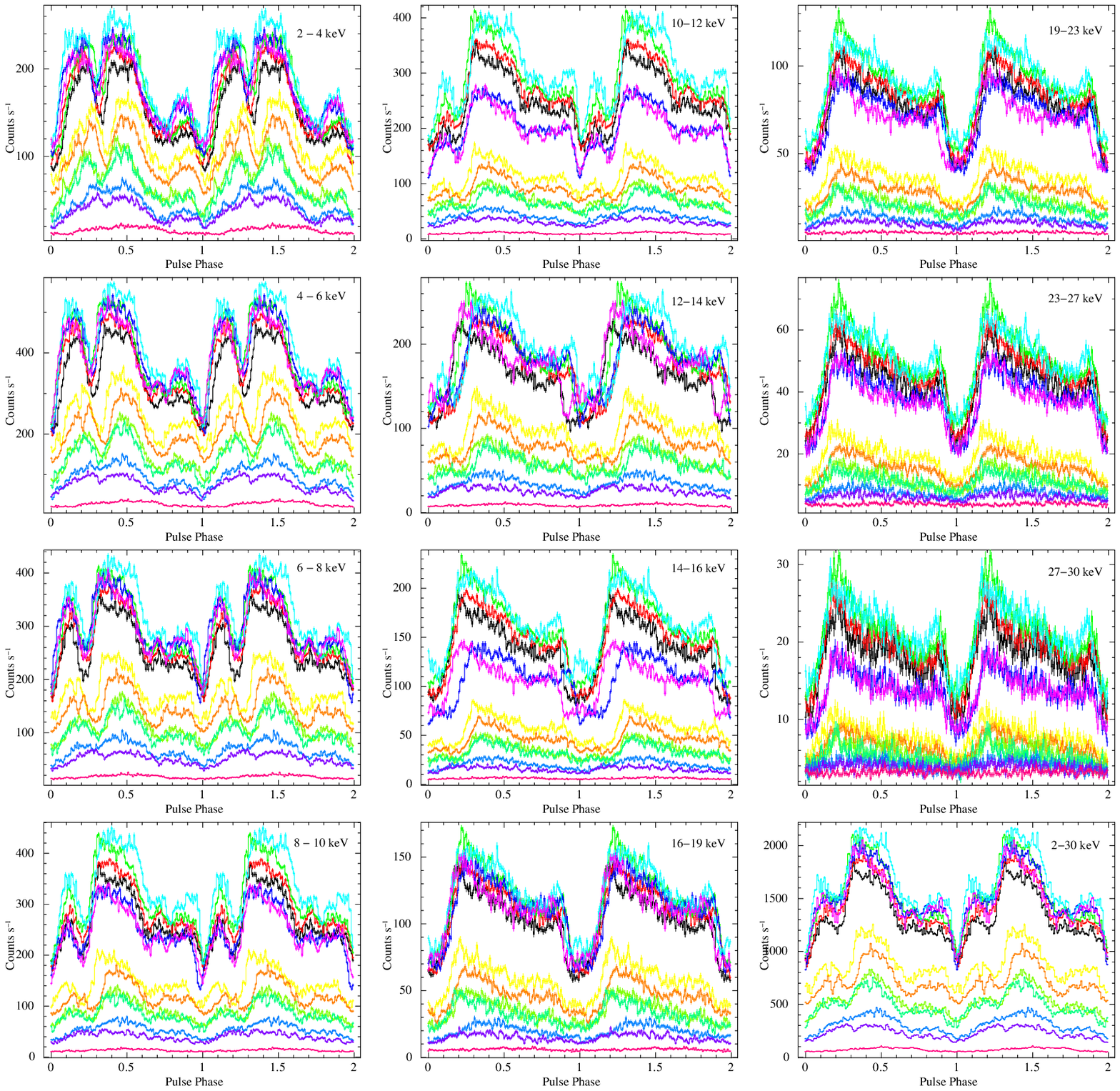}
\vspace{-8 cm}
\caption{Evolution of the pulse profiles of GX 304-1 in diferent energy bands created using all the {\em RXTE}-PCA observations. Light blue=19Aug, Blue=18Aug, Green=15Aug, Magenta=20Aug, Red=14Aug, Black=13Aug, Yellow=23Aug, Orange=24Aug, Yellow+Green=25Aug, Green+Cyan=26Aug, Blue+Cyan=27Aug, Blue+Magenta=28 Aug, Red+Magenta=30 Aug}
\label{all_pp}
\end{figure*}

To study the intensity dependence of the pulse profiles of GX 304-1, we extracted light curves in the 2-60 keV energy band with a time resolution of 0.125 s from Standard1 mode data. The corresponding background count rates were simulated using appropriate background models provided by the {\em RXTE} guest observer facility (GOF). We have chosen good time intervals from each of the observations by filtering the raw data from the effects of satellite pointing offset and times of Earth occultations with standard criteria of offset $\leq$ 0.02 degrees and elevation angle $\geq$ 10 degrees. The background count rates were then subtracted from light curves and the light curves were converted to the solar system barycenter. The outburst  lasting for $\sim$25 days, which occured during August 2010 is observed to have a relatively large intensity compared to the previous outbursts. The peak intensity per PCU is noted to be $\sim$2000 counts s$^{-1}$. The $\sim$ 272 s pulsations are clearly seen in the light curves and just after the peak of the outburst a few light curves show $\sim$2000 s modulation along with the pulsations. This lasted only for a few days. The outburst started around MJD 55410 reached the peak at MJD 55422, then decayed slowly and ended around MJD 55440. The {\em RXTE}-PCA observations started just before the peak (MJD 55421) and the rest of the outburst was well covered. For each day over the outburst, we have searched for pulsations, applying $\chi^{2}$ maximization technique and the best period obtained was used for folding the light curves. GX 304-1 is a slow pulsar and hence the pulse period cannot be determined accurately using just a few short duration observations. The most accurate period we obtain during this outburst is 275.37s on MJD 55423. The pulse periods that we have measured with {\em RXTE}-PCA data and used to create the respective pulse profiles are consistent with the one measured with the {\em Fermi}-GBM $(www.batse.msfc.nasa.gov/gbm/science/pulsars/)$.

The pulse period that we have obtained during this outburst indicates that during the period of quiescence, the pulsar has spun down. The folded light curves for each day covering the 2010 outburst are shown in Figure ~\ref{fig2}. The pulse profiles show significant evolution, as the outburst progresses$:$ complex structures with multiple components were seen on some days. At the start of the outburst, the pulse profile consists of a main peak with steps on both sides $-$ like a shoulder $-$ and a narrow minimum. As the outburst progresses, the depth of the narrow minimum increases. During the decay of the outburst, the pulse profiles become complex with small spikes and the main peak gets broader. At the end of the outburst, the steps on both sides of the main peak disappear resulting in a sinusoid shaped profile with spikes or flares. In the intermediate period, on August 23 and 24, the pulse profile showed a very narrow dip at a phase 0.25 before the main peak, while on 28 August a similar narrow dip was seen coinciding with the main peak.

We have investigated the energy dependence of the pulse profiles for GX 304-1 and its evolution during the outburst.
We extracted 0.125 s light curves from the generic event and generic binned mode data by selecting counts detected with PCU2 detector. Twelve different light curves in different narrow energy bands covering the 2-30 keV range were created from all the observations. The energy dependence of the pulse profiles of GX 304-1 at the peak of the outburst (MJD 55423, August 15) are shown in Figure ~\ref{fig3}. The pulse profiles exhibit significant variation with energy. At lower energies, the pulse profiles consists of multiple peaks, a primary peak at phase 0.5 and a secondary peak at phase 0.2 and a small hump at phase 0.9. With increasing energy, the amplitude of the secondary peak decreases and above 12 keV it completely disappears leaving a broad dip at around phase 0.2. At higher energies, the pulse profile is characterised by a single asymmetric peak with a slight phase shift and broad dip. The maximum of the pulse peak at higher energies appears at the phase at which the second peak occurs at lower energies. The dip has energy dependence, both in phase and in width. The evolution of the pulse profile during the entire outburst is shown in Figure ~\ref{all_pp} for different narrow energy bands. In all the energy bands, the pulse profile is more complex at higher intensity level. Near the end of the outburst the pulse profile is single peaked and is almost identical in all the energy bands. To study the complex energy dependence of the pulse profiles we have carried out pulse phase resolved spectral analysis which is described in section 4. \\

\subsection{Power density spectrum}

\begin{figure}
 \centering
 \includegraphics[width=5.5 cm, angle=-90]{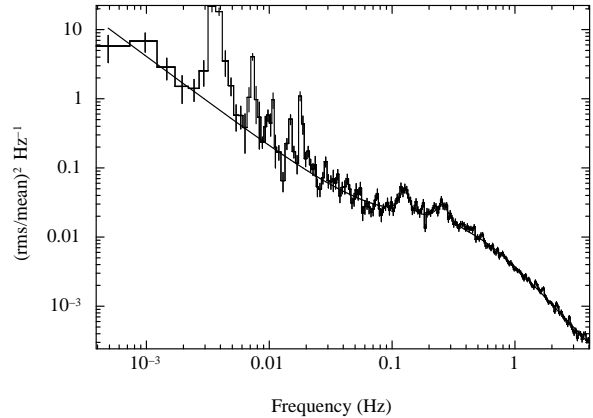}
 \caption{A representative power density spectrum with a QPO feature at 0.125 Hz}
 \label{PDS}
 \end{figure}

QPOs are detected in many high magnetic field accretion powered pulsars which include LMXBs and HMXBs. Aperiodic variabilities like QPOs appear as broad peaks in the power density spectrum. To detect any possible QPO feature in GX 304-1, timing analysis was performed on light curves in the 2-60 keV band  extracted from Standard1 mode data of the PCA with a time resolution of 0.125 s. We generated Power Density Spectra (PDS) using the FTOOL {\em powspec}. Light curves were broken into segments of duration of 2048 s and the PDS of 4-6 segments were averaged to improve the detectability of any QPO like feature. The PDS were normalized such that their integral gives the squared rms fractional variability and the white noise level was subtracted. A narrow peak at around $0.0036$ Hz corresponding to the spin frequency of the pulsar and several harmonics are seen in the PDS. In addition to the spin frequency and its harmonics, two broad QPO features, the first one at 0.125 $\pm 0.002$ Hz are seen in the PDS (Figure ~\ref{PDS}). The second broad feature has a center frequency of 0.257 Hz and is likely to be harmonic of the first. This is the second source after MAXI J1409-619 \citep{Kaur2010} in which harmonics of QPOs are being reported in any accretion powered pulsar. Continuum of the power density spectrum was fitted with two components, a power-law and a Lorentzian and the two QPO features were fitted with additional Lorentzian components (see Figure ~\ref{PDS}). The pulsations and its harmonics are avoided while fitting the continuum.  The background corrected rms value of the QPO feature is low, about 2.9\%. The quality factor of the 0.125 Hz QPO feature is about 7 and the detection significance is 6 $\sigma$.

\begin{figure}
 \centering
 \includegraphics[width=5.5 cm, angle=-90]{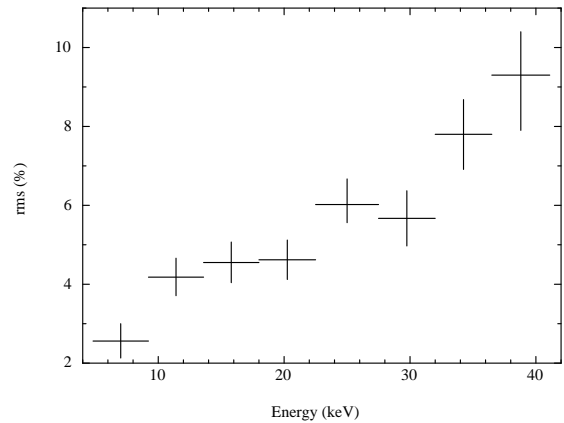}
 \caption{QPO rms variation with energy}
 \label{rms-e}
 \end{figure}

\begin{figure}
 \centering
 \includegraphics[width=5.5 cm, angle=-90]{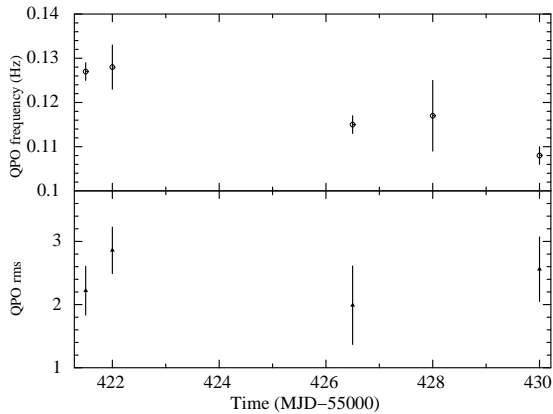}
 \caption{Frequency and rms evolution of the 0.12 Hz QPO feature with time}
 \label{time-fre}
 \end{figure}

Analysis of the event mode data showed that the QPO feature was present in hard X-rays also. To investigate the energy dependence of the QPO feature, we created power density spectra in eight different energy bands and found that the rms value is increasing with the energy. The rms variation in the QPO increases from 2.9\% at 7 keV  to 9\% at 40 keV which is shown in Figure ~\ref{rms-e}. In GX 304--1, the QPO feature is seen up to 40 keV. In accretion powered pulsars, the highest energy up to which QPOs are detected is in the range of 10-25 keV {\citep{Paul&Rao1998,Mukherjee2006,Raichur2008,James2010,James2011,Devasia2011}}, 3A 0535--262 being an exception in which QPOs have been detected only in the hard X-ray band of $>$ 25 keV {\citep{Finger2009}}. We calculated the QPO frequency at different times and the time evolution seen is shown in Figure ~\ref{time-fre}. The QPO frequency decreased from 0.128 Hz to 0.108 Hz in 12 days. \\

\section{Spectral Analysis}
\subsection{Pulse phase-averaged spectroscopy}
We have carried out spectral analysis in the 3-30 keV band on all days during the outburst. The source and background spectra were extracted from Standard2 mode data of the PCA detectors.
The background spectrum was simulated using the {\em pcabackest} tool with appropriate background models provided by the {\em RXTE} guest observer facility (GOF). To fit the 3-30 keV energy spectrum of GX 304-1, we first tried a  model consisting of an absorbed power-law with a high energy cut-off and a Gaussian component for iron fluorescence line emission for all of these spectra. This simple model fits the energy spectrum over the outburst with a reduced $\chi^{2}$ in the range 0.74$-$2.24. For some days, like during the peak of the outburst, the $\chi^{2}$ value is relatively larger. To get a clearer idea, we performed pulse phase resolved spectral analysis for two of the relatively long and bright observations and found that for some phase intervals, this model does not provide a good fit, with a large reduced $\chi^{2}$ value of as much as 10. This reflects the complex nature of the energy dependence of the  pulse profiles shown in Figure ~\ref{fig3} and can be due to the presence of two different spectral components prominent at different energies. We find that the 3-30 keV spectrum of GX 304-1 during most of the outburst can be well fitted by a model consisting of a partial covering power-law  with a high energy cut-off and an iron fluorescent line emission. With this model, we could also fit the phase resolved spectra with better reduced $\chi^{2}$ at all the pulse phases. The iron emission line is found to be quite weak, with equivalent width always less than 100 eV. Since for such weak emission line, the line parameters cannot be measured accurately with the PCA, we have fixed the line centre energy to 6.4 keV and fixed the line width to 100 eV as the iron emission line is known to be narrow in accretion powered pulsars. \\

The analytical form of the model describing the 3-30 keV spectrum of GX 304-1 is given by,

\begin{equation}
N(E) = e^ {-\sigma(E)N_{H1}}(S_{1}+S_{2}e^{-\sigma(E)N_{H2}}+G) E^{(-\Gamma)}I(E)
\end{equation}
where,
\begin{eqnarray}
\nonumber  I(E) & = & 1  \hspace{0.83in} for ~E < E_\mathrm c \\
\nonumber       & = & e^{- \left({E-E_\mathrm c}\over{E_\mathrm f}\right)} \hspace{0.27in}  for ~E > E_\mathrm c
\end{eqnarray}
N(E) is the observed intensity, $N_{H1}$ and  $N_{\mathrm H2}$ are the two equivalent hydrogen column densities, $\Gamma$ is the photon index,  $\sigma$ 
is the photo-electric cross-section, $S_{1}$ and $S_{2}$ corresponds to the respective power-law normalizations, $E_{\mathrm c}$ is the cut--off energy and $E_{\mathrm f}$, the e--folding energy.\\

or, in XSPEC notation$:$ wabs$*$(pcfabs$*$po$*$highec$+$ga)\\

\begin{figure}
\centering
\includegraphics[width=5.5 cm, angle=-90]{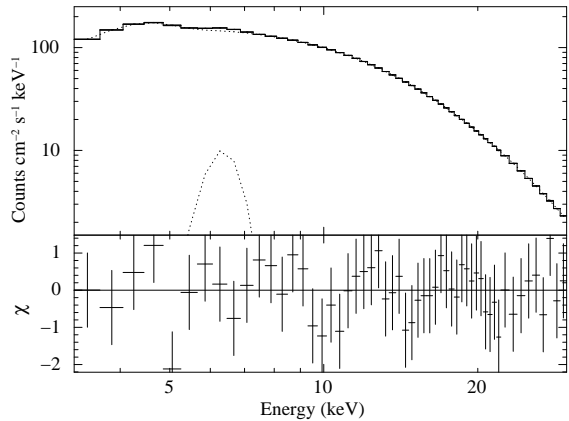}
\caption{The 3-30 keV X-ray spectrum of GX 304-1 obtained with RXTE-PCA on August 15, 2010 (MJD 55423) is shown here along with the best fitted spectral model and the fit residuals.}
\label{peakspec}
\end{figure}

\begin{figure*}
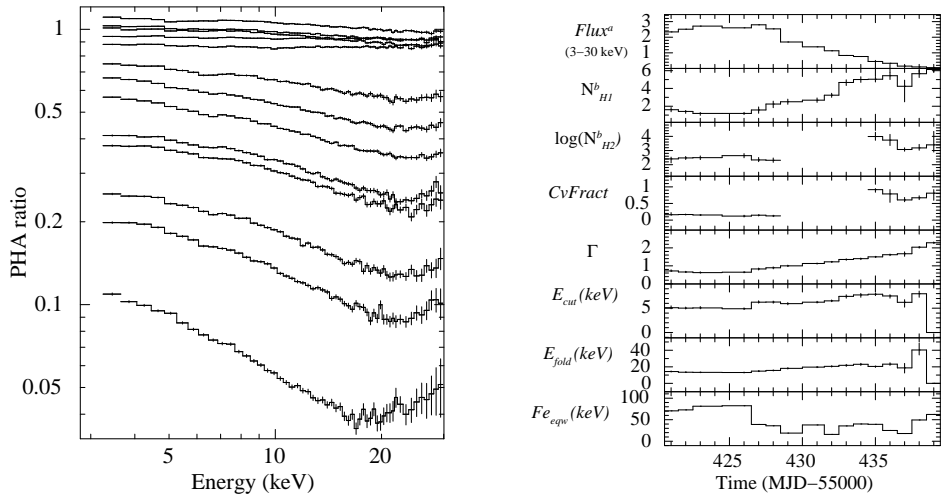

\includegraphics[width=6.5 cm, angle=-90]{fig9_a.ps}\hspace{0.4 in}
\includegraphics[width=6.5 cm, angle=-90]{fig9_b.ps}
\caption{Ratios of the 3-30 keV X-ray spectrum obtained on different days of observations with the spectrum obtained on August 15 are shown in the left panel. The curves correspond to respective days from top: Aug19, Aug18, Aug20, Aug14, Aug13, Aug21, Aug23, Aug24, Aug25, Aug26, Aug27, Aug28, Aug29, Aug30, Aug31. A softening of the spectrum below about 18 keV and a hardening above 18 keV is evident during the decay of the outburst. Evolution of the spectral parameters during the outburst is shown in the right panel.}   
$ ^a ${$10^{-8}$ ergs cm$^{-2}$ s$^{-1}$ for 3-30 keV}\\
$ ^b ${10$^{22}$ atoms cm$^{-2}$}\\
\label{ratiospec}
\end{figure*}

One representative spectrum (obtained on 15th August, MJD 55423) along with the best fitted model and the corresponding residuals are shown in Figure ~\ref{peakspec}. The source showed a significant spectral evolution during the outburst. To illustrate this, ratios of the 3-30 keV X-ray spectrum obtained on different days of observations with the spectrum obtained on August 15 are shown in the left panel of Figure ~\ref{ratiospec}. A softening of the spectrum below about 18 keV and a hardening above 18 keV is evident during the decay of the outburst. During most of the outburst, the 3-30 keV spectrum of GX 304-1 can be well fitted with a partial covering power-law model with a high energy cutoff and iron fluorescent line emission. For a few of the observations carried out during the decay of the outburst, the partial covering absorption component is found to change to single component absorption. Evolution of the spectral parameters during the outburst is shown in the right panel of Figure ~\ref{ratiospec}. The errors given here are for the 1 $\sigma$ confidence level determined using the XSPEC {\em error} comand. \\

 \subsection{Pulse phase-resolved spectroscopy}
\begin{figure*}
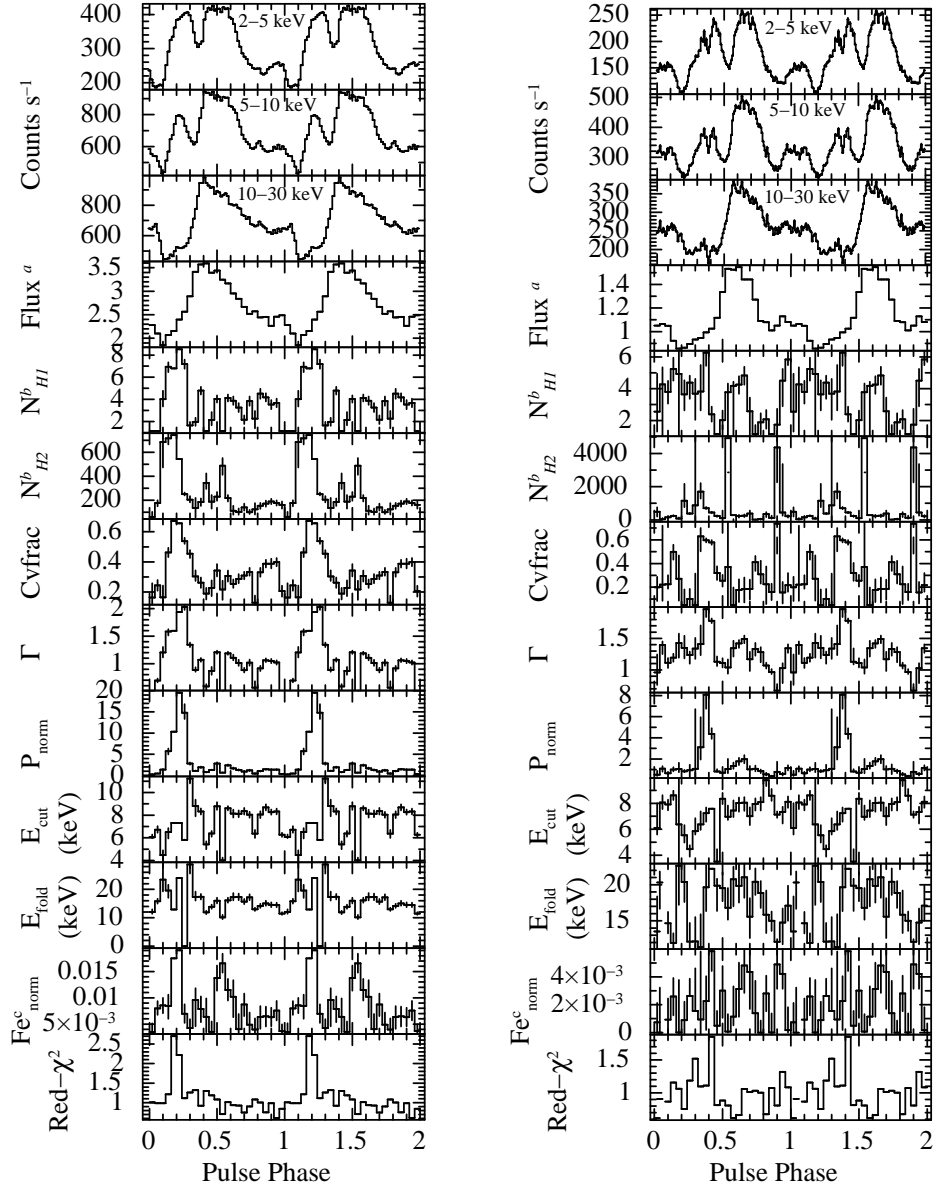

\includegraphics[width=16 cm,  angle=-90]{fig10_a.ps}\hspace{0.2in}
\includegraphics[width=16 cm, angle=-90]{fig10_b.ps}
\caption{Variation of spectral parameters with phase is shown for two observations, one near the peak of the outburst (15 August- MJD 55423) in the left panel and one in the middle of the outburst (August 24- MJD 55432) in the right panel.} 
$ ^a ${$10^{-8}$ ergs cm$^{-2}$ s$^{-1}$ for 3-30 keV} \\
$ ^b ${10$^{22}$ atoms cm$^{-2}$}  \\
$ ^c $photons cm$^{-2}$ s$^{-1}$ \\
\label{ppr}
\end{figure*}

\begin{figure*}
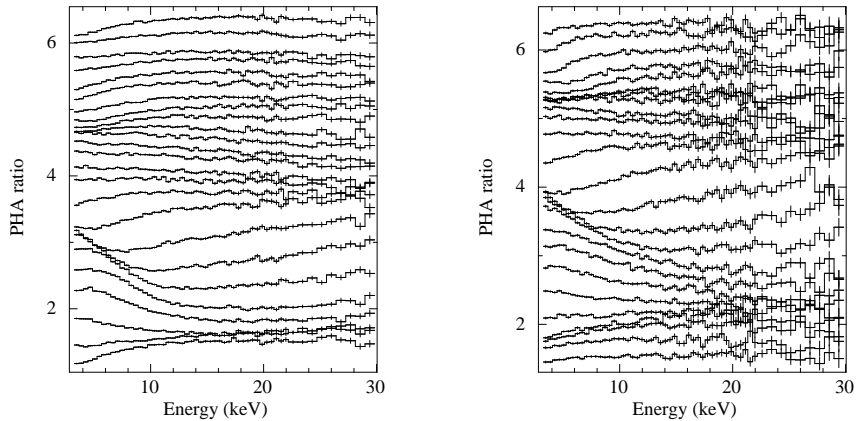

\centering
\includegraphics[width=5.5 cm, angle=-90]{fig11_a.ps}\hspace{0.4in}
\includegraphics[width=5.5 cm, angle=-90]{fig11_b.ps}
\caption{The ratio of the spectra at different pulse phases with the phase averaged spectrum are shown here for the same observations (leftpanel: August 15- MJD 55423 and right panel: August 24- MJD 55432) as in Figure \ref{ppr}.}
\label{pprratio}
\end{figure*}

As mentioned in section 3 and shown in Figures 3 and 4, the pulse shape of GX 304-1 changes dramatically with energy. To have a better understanding of how the energy spectrum changes as the highly magnetised pulsar spins, we performed pulse phase resolved spectral analysis.
A strong energy dependence of the pulse profile is detected till 26th August (MJD 55434). The pattern of energy dependence is similar during 13 August to 26th August. Not all the observations have good enough statistics (count rate $\times$ observation duration) to carry out pulse phase resolved spectroscopy. We have selected two long observations for phase resolved spectroscopy, one near the peak of the outburst (15th August) and one in the middle of the decay (24th August). The data were divided into 25 pulse phase bins and from each bin we have created source energy spectra using the ftool {\em saextrct}. Appropriate background models were used for simulating the background and the response matrix was created using the ftool {\em pcarsp}. Spectral fits were carried out using the model described in eqn(1) and we have varied all parameters except the center and width of the Gaussian line component. The variations of the free parameters with pulse phase are shown in Figure \ref{ppr}. The partial covering parameters show significant variation with pulse phase. The spectrum also shows softening and higher normalisation in certain pulse phases. We would like to mention here that the variations in some of the spectral parameters, like the photon index, cut-off energy and folding energy are not completely independent. However, we emphasise the fact that there is a large increase in the column density of the partial covering component in certain pulse phases. The ratio of the spectra at different pulse phases with the phase averaged spectrum are shown in Figure \ref{pprratio} for both the observations. The strong pulse phase dependence of the spectrum at energies below 15 keV is very clear in the Figure. We also note here that in some pulse phases the phase resolved spectra do give an acceptable fit to a spectral model without partial covering. These phases are represented with small N$_{\rm H2}$ in Figure \ref{ppr}. However, in the remaining phases, we are left with a rather high reduced $\chi^2$ of up to 10. However, our main emphasis is on a large increase in the column density of the partial covering component in certain pulse phases.


\section{Discussion}

Be/X-ray binaries are observed to be transient sources in X-rays exhibiting recurrent outbursts which imply episodic accretion of material during the course of its evolution. It is believed that most of the accretion is from the Be star's circumstellar disc, as the neutron star passes through the radially extended equatorial disc. Depending upon the mode of accretion and the size and orientation of the disc with respect to the neutron star, the pattern of X-ray outbursts may change in relative strength and duration, which in turn allows detailed probing of the accretion process over a range of mass accretion rates. So transient sources are excellent candidates to study intensity variations over time in the X-ray regime.

 If the accreted material has sufficient angular momentum, an accretion disc is formed around the neutron star and the material spirals inwards. Due to the presence of the high magnetic field of the neutron star, regardless of the existence of an accretion disk, the infalling material reaching the magnetosphere gets channeled on to the magnetic poles forming an accretion stream. Depending on the mass accretion rate, the size of the hot spot, the density of the accretion column and hence the radiation pressure near the surface of the neutron star varies, which in turn results in highly anisotropic emission from the magnetic poles. For a distant observer, the viewing angle changes constantly and hence the resulting spectral parameters vary with phase. The observed X-ray modulation will also be affected by height dependent relativistic light bending around the neutron star \citep{Leahy1995, Riffert1993}. If any inhomogeneities are present in the accretion disc, it is reflected as quasi-periodic oscillations in the power spectrum. The detection of quasi-periodic oscillations helps us to infer the magnetic field of the neutron star. In the previous section we have presented results from an extensive analysis of periodic and aperiodic timing properties, energy resolved timing, intensity dependent and pulse phase dependent spectral properties of the transient X-ray pulsar GX 304-1 over a large range of X-ray intensity during an outburst. Here we discuss possible implications of some of the key observed features.

\subsection{Quasi periodic oscillations}

GX 304-1 is the 22nd high magnetic field accretion powered pulsar in which QPOs have been detected; this includes the recent discovery of 0.09 Hz QPO in 1A 1118-61 \citep{Devasia2011}. Of the 22  pulsars in which QPOs have been detected, 15 of them are in HMXBs including this new discovery. \\

Two widely accepted models which can explain the QPO production mechanism are the Keplerian Frequency Model (KFM) and the Beat Frequency Model (BFM). In the former model, inhomogeneous structures  in the inner edge of accretion disk rotate at the keplerian frequency, which modulate the X-ray flux by obscuration, at $\nu_{k}$, resulting in a QPO at $\nu_{QPO}= \nu_{k}$ \citep{Klis1987}. Blobs of matter in the inner accretion disk rotate at the keplerian frequency, while the the magnetic field lines rotate at the neutron star spin frequency. Because of the coupling between the accretion disk  and the neutron star through the magnetic field lines, the mass flow to the neutron star is modulated at the beat frequency between  $\nu_{k}$ and $\nu_{s}$ resulting in a QPO. So the latter model  suggests that QPOs occur at $\nu_{QPO}= \nu_{k} -\nu_{s}$ \citep{Alpar&Shaham1985}.

The QPO features in X-ray pulsars may be transient or persistent, independent of the transient or persistent nature of source. Most of the QPO features discovered are transient in nature. In KS 1947+300 and 4U 1901+03 the QPO feature was seen  during the end of the outburst \citep{James2010, James2011}. But a few sources like 4U 1626--67 \citep{Kaur2008}, XTE J1858+034, \citep{Paul&Rao1998, Mukherjee2006} and 1A 1118--61\citep{Devasia2011},  show QPOs most of the time. In the source GX 304--1, the 0.12 Hz QPO feature is very similar to that in 1A 1118-61 and XTE J1858+034. In all these sources the QPO frequency changes with time. In GX 304--1, the QPO frequency changed from 0.128 Hz to 0.108 Hz in 12 days. One peculiar feature of this QPO is that it is seen up to 40 keV unlike all other QPO sources. The rms  value shows a positive energy dependence as in KS 1947+300 \citep{James2010}. \\
 
In the 272 s pulsar GX 304--1, because the QPO frequency is two hundred times larger than the spin frequency it cannot distinguish between KFM or BFM; therefore the QPO could be explained by either mechanism. The radius of the QPO production region can be calculated as, 

\begin{equation}
R_{qpo} = \left({GM}\over{4 \pi^{2} \nu^{2}_{k}}\right)^{1/3}
\end{equation}

Assuming a neutron star mass of 1.4 $M_{\odot}$, the inner radius of the accretion disk is obtained to be about $6.2\times 10^{3}$ km. The highest 2-30 keV X-ray flux during the time of QPO detection is 5.8$\times$10$^{-9}$ erg cm$^{-2}$ sec $^{-1}$ and the corresponding QPO frequency is 0.128 Hz.  The total X-ray luminosity for the  source is calculated to be about $0.282\times10^{37}$ erg s$^{-1}$ at 2 kpc. The magnetospheric radius of a neutron star can be expressed in terms of the luminosity and magnetic moment as \citep{Frank&Raine2002},

\begin{equation}
R_{m}= 3\times10^{8}L^{-2/7}_{37}\mu^{4/7}_{30}  ~\\ cm
\end{equation}

where $L_{37}$ is the X-ray luminosity in units of 10$^{37}$ erg s$^{-1}$ and $\mu_{30}$ is the magnetic moment in units of 10$^{30}$ Gauss cm$^{3}$. Equating $R_{m}$  with $R_{qpo}$ we estimate the magnetic moment  as 2.11$\times$10$^{30}$ Gauss cm$^{3}$, i.e magnetic field strength in the range 2.1 to 4.2 $\times$10$^{12}$ Gauss depending on magnetic latitude. A similar calculation using the lowest X-ray flux ($0.185\times10^{37}$ erg s$^{-1}$) during which QPOs are detected and the corresponding QPO frequency (0.108 Hz) gives a magnetic field strength of the order of 2.0$\times$10$^{12}$ Gauss. The cyclotron line detected in this source occurs at 51 keV, implying a magnetic field strength  of around  4.2 $\times$10$^{12}$ Gauss, comparable  with the magnetic field strength obtained from the QPO frequency and X-ray luminosity measurements.

\subsection{Pulse profile evolution and energy dependence}

The pulse profiles of accretion powered pulsars show dramatic variations with intensity as well as energy both in transient and persistent sources \citep{White1983, Nagase1989, Raichur2010}. Many geometrical and physical processes occuring near to the neutron star can affect the pulse profiles. Variable mass accretion rates during the outburst may result in the evolution of the pulse profile with changes in density and structure of the accretion column. Strong spectral and geometrical variations in the accretion columns may cause significant variations in the shapes of the pulse profiles. The plasma layer flowing along the Alfven surface spinning with the neutron star and covering a part of the neutron star surface will regularly shield the emission region from the observer' s line of sight. The optical depth of the various parts of this layer changes with time and may result in the observed changes with intensity and energy \citep{Basko1976}. Radiation supported scattering atmospheres, formed surrounding the accretion column can also lead to energy dependent features in the pulse profiles \citep{Sturner1994}. Transient pulsars like EXO 2030+375, V0332+53 \& 4U 0115+63 \citep{Parmar1989, Tsygankov2007} having a Be star companion exhibit strong variations in pulse profiles during the outburst. In some cases, the pulse profiles are relatively complex with the profile during the peak of the outburst different from the profiles during the decay phases e.g: 1A 1118-61, another slow pulsar showing similar variations in the pulse profiles \citep{Devasia2011}. While there is no straightforward interpretation of pulse profile changes with time and energy, we find that near the end of the decay of the outburst, the pulse profile has  a single broad peak and less energy dependence. In contrast, during the peak and first half of the decay of the outburst, the pulse profile is complex, with greater level of complexity at lower energy. Pulse phase resolved spectroscopy indicates that the strong energy dependence of the pulse profile can be due to a combination of pulse phase dependence in the underlying continuum emission and pulse phase locked partial covering.

\subsection{Pulse phase-averaged and pulse phase-resolved spectra}
The energy spectra of most  accreting X-ray pulsars can be well described by a simple model consisting of an absorbed power-law with a high-energy cut-off (around 10$-$20 keV in most cases) and a Gaussian component for the iron line fluorescence emission. The X-ray spectrum of GX 304-1 does not give an acceptable fit with this simple model. Instead it requires a partial covering model which fits the source spectrum quite well. In this case, a part of the emission from the source is blocked from the observers line-of-sight by material in the accretion column that is spinning with the neutron star.

The ratio of the pulse phase averaged spectrum obtained on different days of the outburst with the same during the peak of the outburst (left panel of Figure 9) shows a significant spectral evolution. As the outburst decayed, the spectrum became softer up to about 15 keV and there was a possible hardening of the spectrum above 15 keV. The spectral parameters measured from these spectra (right panel of Figure 9) show a corresponding gradual increase in the spectral index. Emergence of a hard component above 15 keV during the decay of the outburst shows a corresponding increase in the covering fraction and the column density of the partial covering component. 
Becker and Wolff (2007) computed the expected X-ray spectrum of accretion powered high magnetic field pulsars from bulk and thermal Comptonization of seed photons in the accretion column. The pulse phase averaged and accretion column height integrated X-ray spectrum that emerges from such sources have a power-law shape with a high energy cut-off as is seen in GX 304-1. The computed X-ray spectrum has a low energy turn-over also (Becker \& Wolff 2005), which in empirical spectral modeling is often ascribed to an absorption column density. The {\em RXTE}-PCA data presented here does not include this energy band below 3 keV. With decreasing X-ray luminosity, the X-ray spectrum may also show a softening, as is seen in GX 304-1. However, these effects have not yet been quantified and so a direct comparison with the measured spectra cannot be made.

Pulse phase resolved spectral analysis shows a strong variation in the partial covering parameters over  phase, supporting the scenario. Similar properties have  been seen in a few other HMXBs like 1A 1118-61 \citep{Devasia2011} and GRO J1008-67 \citep{Naik2011}. In these sources, a large increase in the column density of the partial covering component is compatible with a scenario in which the accretion stream from the inner accretion disk to the pulsar is narrow and phase locked, causing a partial absorption of the emitting region in certain pulse phases. While the scenario presented here for GX 304-1 and elsewhere for 1A 1118-61 and  GRO J1008-57 are entirely consistent, a rather high value of N$_{H2}$ is found in certain pulse phases. It cannot be ruled out that the observed spectral variation and pulse phase dependence can also be modeled with multiple hard X-ray emission components with independent pulse phase dependencies, provided consistent with the pulsar emission models (Becker \& Wolff 2007).

\section{Conclusions}
We have carried out an extensive timing and spectral analysis of {\em RXTE}-PCA observations of the transient X-ray pulsar GX 304-1 during an outburst in 2010. The key results obtained are
\begin{itemize}
\item Detection of QPOs, along with a harmonic which is rare in X-ray pulsars. The magnetic field estimated using the QPO frequency and the X-ray luminosity is comparable to the one estimated from the cyclotron absorption feature in this source.
\item The pulse profile varied from a complex shape during the outburst to a simple profile near the end of the outburst. During the bright state, the pulse profile had a complex energy dependence, with multiple peaks in the low energy band and a single peak with an absorption dip in the high energy band. This indicates the presence of a source spectrum with multiple, pulse phase dependent components.
\item The X-ray spectrum softened as the outburst decayed, and showed presence of strong partial covering absorption on some days.
\item Pulse phase resolved spectroscopy, carried out on some observations showed large absorption column density in some pulse phases, which can be caused by absorption in narrow, phase locked accretion streams.
\end{itemize}

\section{Acknowledgements} 
We thank an anonymous referee whose valuable comments helped us to improve the contents of the paper. This research has made use of data obtained through the High Energy Astrophysics Science Archive Research Center Online Service, provided by the NASA/Goddard Space Flight Center.

--------------------------------
\def\etal{{\it et~al.\ }}
\def\apj{{Astroph.\@ J.\ }}
\def\apjl{{Astroph.\@ J. \@ Lett. }}
\def\araa{{Ann. \@ Rev. \@ Astron. \@ Astroph.\ }}
\def\mn{{Mon.\@ Not.\@ Roy.\@ Ast.\@ Soc.\ }}
\def\aap{{Astron.\@ Astrophys.\ }}
\def\aj{{Astron.\@ J.\ }}
\def\prl{{Phys.\@ Rev.\@ Lett.\ }}
\def\pd{{Phys.\@ Rev.\@ D\ }}
\def\nucp{{Nucl.\@ Phys.\ }}
\def\nat{{Nature\ }}
\def\sci{{Science\ }}
\def\plb {{Phys.\@ Lett.\@ B\ }}
\def \jetpl {JETP Lett.}

\end{document}